\documentclass{article}

\usepackage[nonatbib,preprint]{nips_2018}

\usepackage[utf8]{inputenc} 
\usepackage[T1]{fontenc}    
\usepackage{hyperref}       
\usepackage{url}            
\usepackage{booktabs}       
\usepackage{amsfonts}       
\usepackage{nicefrac}       
\usepackage{microtype}      
\usepackage{graphicx}
\usepackage{subcaption}
\usepackage{comment}

\title{Improving Traffic Safety Through Video Analysis in Jakarta, Indonesia}

\author{
    Jo\~ao Caldeira\thanks{J.\ Caldeira, A.\ Fout, A.\ Kesari, and R.\ Sefala contributed equally to this work.} \\
    Department of Physics\\
    University of Chicago\\
    \texttt{jcaldeira@uchicago.edu} \\
    \And
    Alex Fout\footnotemark[1] \\
    Statistics \\
    Colorado State University \\
    \texttt{alex.fout@colostate.edu} \\
    \And
    Aniket Kesari\footnotemark[1] \\
    Jurisprudence \& Social Policy \\
    University of California, Berkeley \\
    \texttt{akesari@berkeley.edu} \\
    \And
    Raesetje Sefala\footnotemark[1] \\
    Machine Learning \\
    University of the Witwatersrand \\
    \texttt{raesetje.sefala@students.wits.ac.za} \\
    \And
    Joseph Walsh \\
    Center for Data Science and Public Policy \\
    University of Chicago \\
    \And
    Katy Dupre \\
    Center for Data Science and Public Policy \\
    University of Chicago
    \And
    Muhammad Rizal Khaefi \\
    Pulse Lab Jakarta \\
    \texttt{muhammad.khaefi@un.or.id}
    \And
    Setiaji \\
    Jakarta Smart City \\
    \texttt{setiaji@jakarta.go.id}
    \And
    George Hodge \\
    Pulse Lab Jakarta \\
    \texttt{george.hodge@un.or.id}
    \And
    Zakiya Aryana Pramestri \\
    Pulse Lab Jakarta
    \And
    Muhammad Adib Imtiyazi \\
    Jakarta Smart City
}

\begin{document}

\maketitle
\setcounter{footnote}{0}

\begin{abstract}
    This project presents the results of a partnership
    between the Data Science for Social Good fellowship,
    Jakarta Smart City and Pulse Lab Jakarta to create a video analysis pipeline for the purpose of improving traffic safety in Jakarta. The pipeline transforms raw traffic video footage into databases that are ready to be used for traffic analysis. By analyzing these patterns, the city of Jakarta will better understand how human behavior and built infrastructure contribute to traffic challenges and safety risks. The results of this work should also be broadly applicable to smart city initiatives around the globe as they improve urban planning and sustainability through data science approaches.\footnote{The source code developed in this project is available at \url{https://github.com/dssg/jakarta_smart_city_traffic_safety_public/}.}
\end{abstract}

\section{Introduction}

    The World Health Organization's \textit{Global status report on road safety 2015} estimates that over 1.2 million people die each year in traffic accidents~\cite{WHO:tNYZ6LfW}. Nearly 2000 such fatalities occur annually in the city of Jakarta, Indonesia, making it one of the most dangerous cities in the world for traffic safety. Many of these deaths are preventable through effective city planning.
 
    Jakarta has experienced rapid population growth over the last 50 years, from roughly two million people in 1970 to more than 10 million today. With this growth comes a rise in vehicle ownership and congestion, and these factors inevitably lead to an increase in the number of traffic incidents. 
    
    One of the core problems with using machine learning and other data-driven techniques in traffic safety analysis is that it is difficult to collect high-quality data. In partnership with Jakarta Smart City (JSC)\footnote{\url{https://smartcity.jakarta.go.id/}} and Pulse Lab Jakarta (PLJ)\footnote{\url{http://pulselabjakarta.org/}}, a team of fellows at the Data Science for Social Good (DSSG) fellowship at the University of Chicago\footnote{\url{https://dssg.uchicago.edu/}} was formed to tackle this problem.
    Our team was given access to video footage of seven traffic cameras spread over Jakarta. We developed a video analysis pipeline that furnishes JSC and PLJ with the ability to generate rich databases that contain massive amounts of information about traffic behaviors. We hope our work provides a roadmap for applying machine learning throughout the developing world in the context of smart cities and urban planning more broadly.
    
    In addition to the technical application of the video analysis pipeline, we want this project to provide a template for others who hope to successfully deploy machine learning and data driven systems in the developing world. Through intense cooperation between the fellowship team in Chicago and the project partners in Jakarta, we gleaned insights into how to effectively build a system that is likely to be used by a partner in the developing world. Specifically, we became attuned to the need for mapping technical solutions to social problems that are articulated by people working in the field, understanding cultural context and awareness, and creating a feasible deployment strategy. These lessons should be invaluable to the many researchers and data scientists who wish to partner with NGOs, governments, and other entities that are working to use machine learning in the developing world. 
    
\section{Methodology}

\subsection{Data}

JSC provided approximately 700GB of 1024 by 768 pixel video footage taken from seven locations across Jakarta, chosen to represent varying geography, infrastructure, and traffic behavior. Additional video footage from other locations was downloaded using JSC's public data portal. Starting from these videos, we were tasked with generating quantitative data that could be used for more standard traffic analysis.

In order to evaluate our results, we needed to obtain annotated videos. This was done by hand-labeling vehicles and pedestrians in a sample of our videos using the \emph{Computer Vision Annotation Tool} (CVAT)~\cite{CVAT:jxyQOWpJ}.

\subsection{Translating Data to Methods}

Before translating raw video into structured data, extensive work had to be done so all partners had a common vision of the policy interventions that the Jakarta authorities hoped to deploy given better traffic information. We established that in the medium term, they are interested in learning the best places that they can place ``traffic stewards'' and build traffic lights. In the long term, they are interested in learning where bigger infrastructure projects may be most successful. 

In addition to these specific interventions, we also set out to define the scope of problematic traffic behaviors that the city hopes to curtail. In this case, we are concerned with a few specific behaviors, including vehicles driving against traffic, motorcycles and scooters driving on pedestrian surfaces like sidewalks, and illegal stopping or parking. Once we understood the most dangerous driving behaviors, and the policy levers available, we were able to think about how to map social policy problems to technical solutions. This map informed the specific data that we generated. We detail our particular choice of computer vision methods in Section~\ref{cv_methods}.
    
\section{Results}

\subsection{Pipeline}
Our pipeline was created with a ``streaming'' approach, which breaks a video into individual frames at the beginning of the pipeline. It then passes these frames through a system of workers and queues. Essentially, each worker is given a particular “task” (e.g.\ object detection) that it performs on each frame. Once it finishes a task, it sends that frame to the next queue, where the frame waits until the next worker is ready to process it. Frame order is preserved, and at the end, a worker puts frames back together to output the original video with any new annotations or analysis. The workers also output quantitative information about object counts, direction, etc.\ which can be loaded to a database. 

The pipeline is modular, so any worker can be replaced by a different algorithm. Modularity is a key feature, allowing a user of the pipeline to optimize its performance on their specific task of interest. This also avoids loading large uncompressed videos into memory as implied by a batched approach, permits simultaneous execution of multiple tasks, and permits load balancing by adding or removing workers from tasks as necessary. There are some apparent limitations to this decision, namely that GPU computations utilized by many machine learning algorithms are optimized for batch computation, and workers cannot use yet-unseen frames when performing a task, which limits the exploitation of temporal dependence between frames. We note that both of these concerns can be addressed through use of appropriate buffering in the workers, which is exactly how we perform efficient classification with YOLOv3 (see below).

\subsection{Detection, Motion, and Segmentation} \label{cv_methods}

\begin{figure}[ht]
    \centering
    \begin{subfigure}[]{0.45\textwidth}
        \includegraphics[width=\textwidth, height = 3cm]{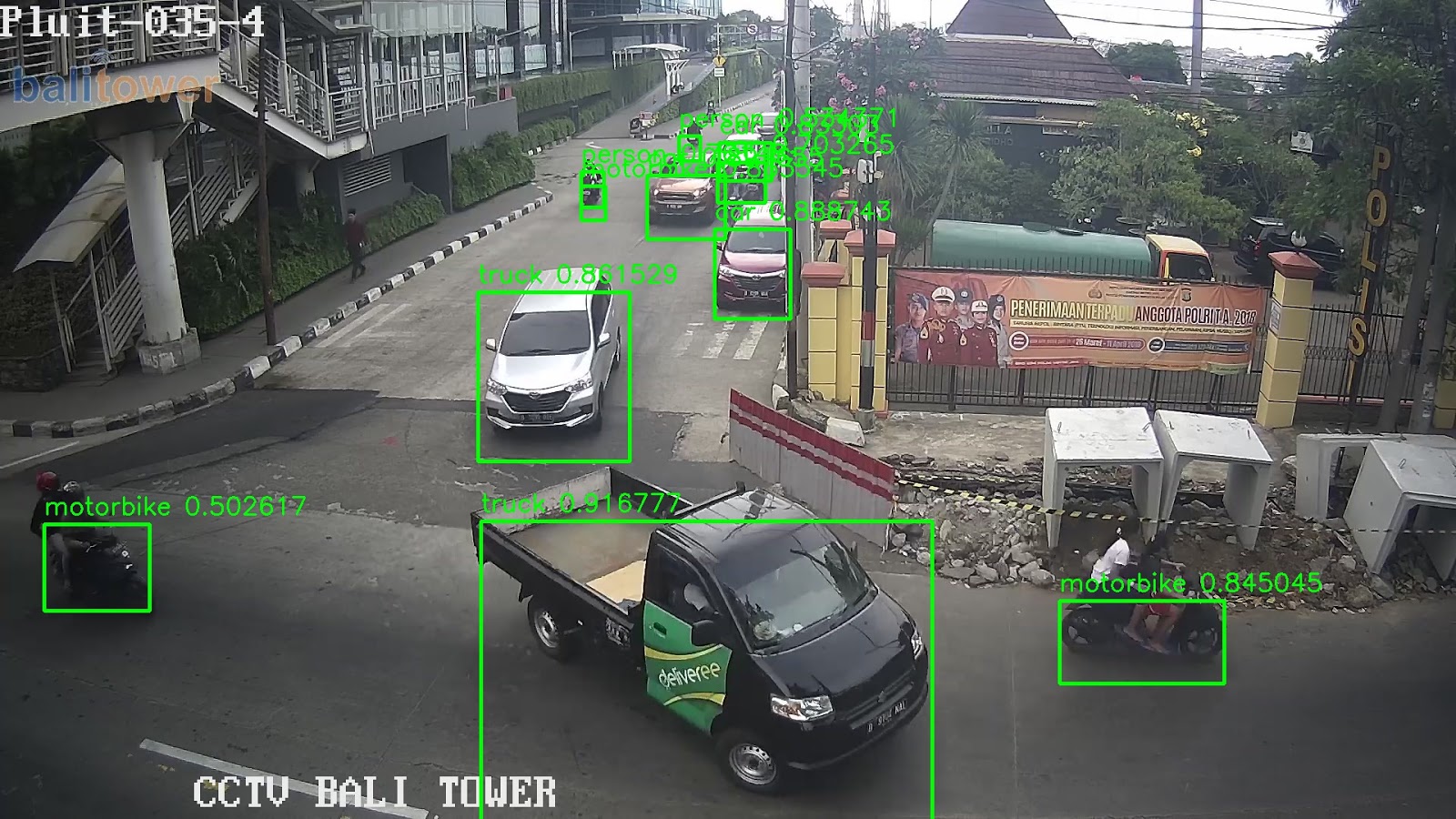}
    \end{subfigure}
    ~ 
    \begin{subfigure}[]{0.45\textwidth}
        \includegraphics[width=\textwidth, height = 3cm]{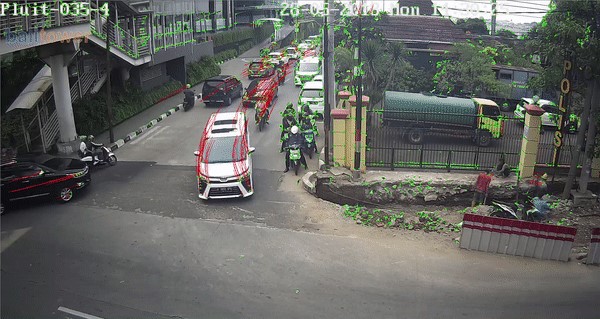}
    \end{subfigure}
    \caption{On the left, detection/classification with YOLOv3. On the right, motion detection with Lucas-Kanade Optical Flow.}
    \label{fig:detect_motion}
\end{figure}

We developed several modules that make up the pipeline, and are directly related to Jakarta's specific policy requirements. Detection and classification are necessary components as they determine what objects in a frame are motor vehicles. The results of detection and classification provide the foundation for detecting specific traffic behaviors. We use YOLOv3 trained on the COCO dataset~\cite{Redmon:2018uv}. Figure~\ref{fig:detect_motion} (left) shows an example of our detection and classification results.

Motion estimation is similarly important because it helps determine when a vehicle is traveling the wrong way. We chose to use optical flow as it allowed us to extract the information of whether an object was moving and in what direction, without any additional training. Using Lucas-Kanade optical flow algorithm~\cite{lucas1981iterative} in conjunction with Shi-Tomasi feature detection~\cite{shi1993good}, we were able to calculate the direction of movement for every detected vehicle in a frame. We used the existing implementation available in OpenCV~\cite{opencv_library}. Figure~\ref{fig:detect_motion} (right) illustrates this result.

\begin{figure}[ht]
    \centering
    \begin{subfigure}[]{0.45\textwidth}
        \includegraphics[width=\textwidth, height = 3cm]{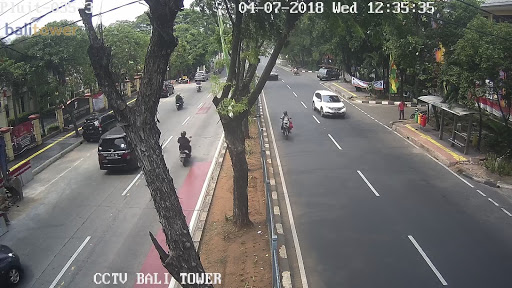}
    \end{subfigure}
    ~ 
    \begin{subfigure}[]{0.45\textwidth}
        \includegraphics[width=\textwidth, height = 3cm]{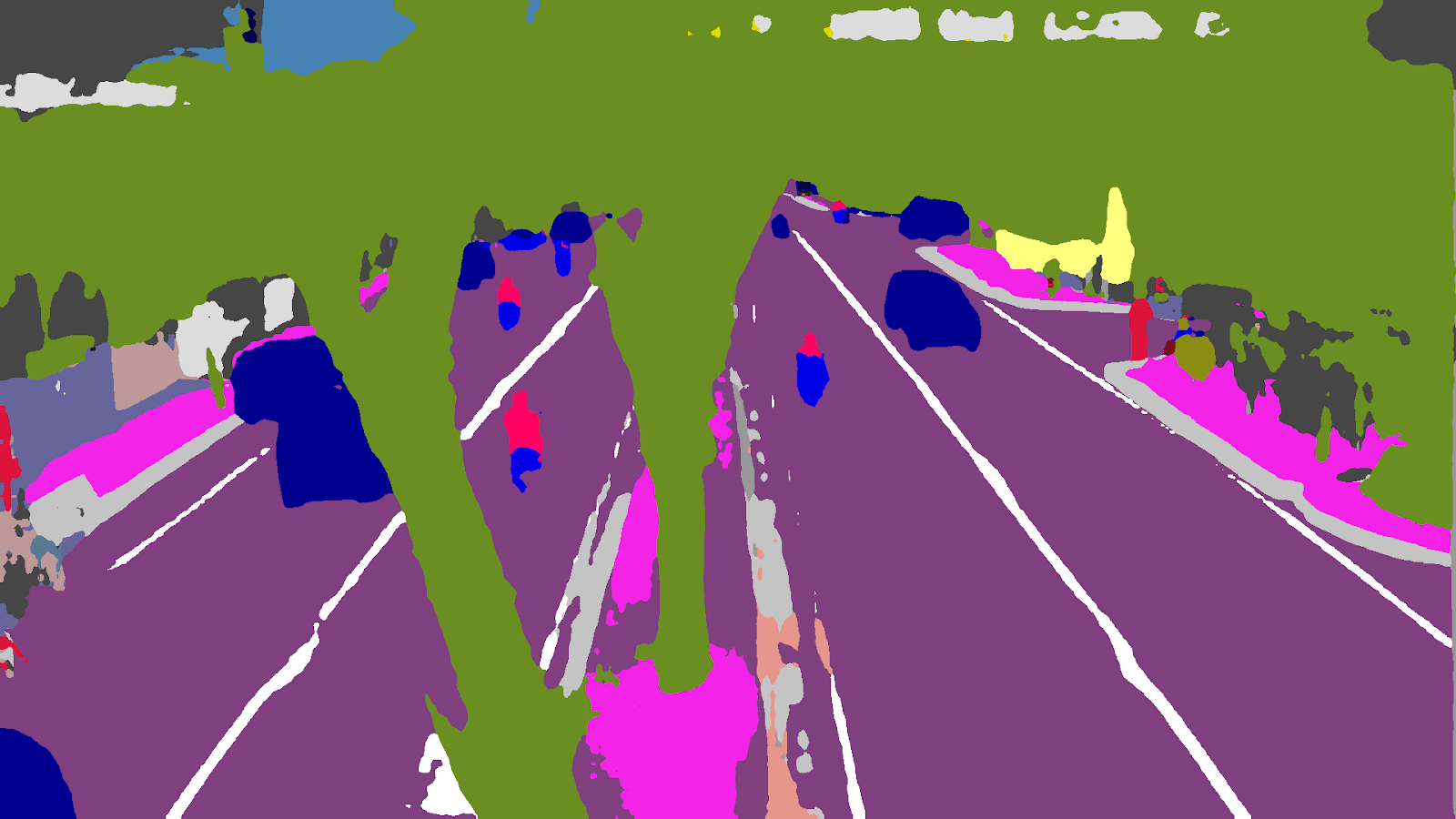}
    \end{subfigure}
    \caption{A scene pre- and post-segmentation.}
    \label{fig:segment}
\end{figure}

Finally, we needed to classify the different regions of the image into different classes, such as road or sidewalk, in order to determine whether motor vehicles were moving in an illegal way. For this task, one can use semantic segmentation, which classifies each pixel as belonging to one of several different classes. We used a pretrained version of the WideResNet-38 model described in~\cite{rotabulo2017place}. The result of semantic segmentation on one of the stretches of road we had data for can be seen in Figure~\ref{fig:segment}. More granular classification of different segments of road, such as encoding the correct direction to drive in or where crosswalks exist, can then be added by hand in each intersection.

\begin{figure}[ht]
    \centering
    \begin{subfigure}[]{0.45\textwidth}
        \includegraphics[width=\textwidth, height = 2.5cm]{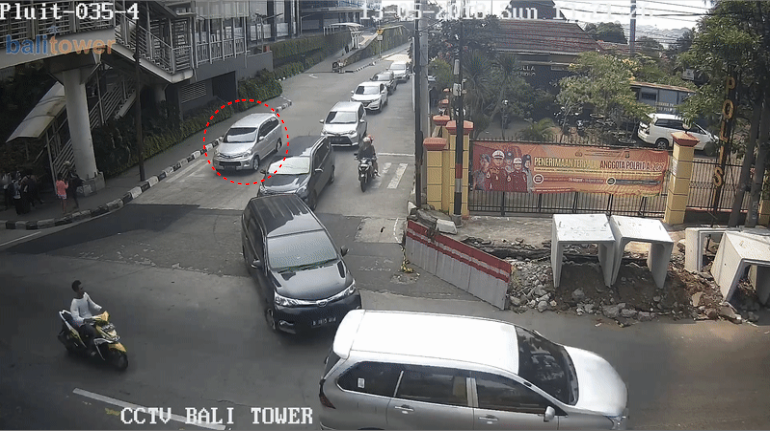}
    \end{subfigure}
    ~ 
    \begin{subfigure}[]{0.45\textwidth}
        \includegraphics[width=\textwidth, height = 2.5cm]{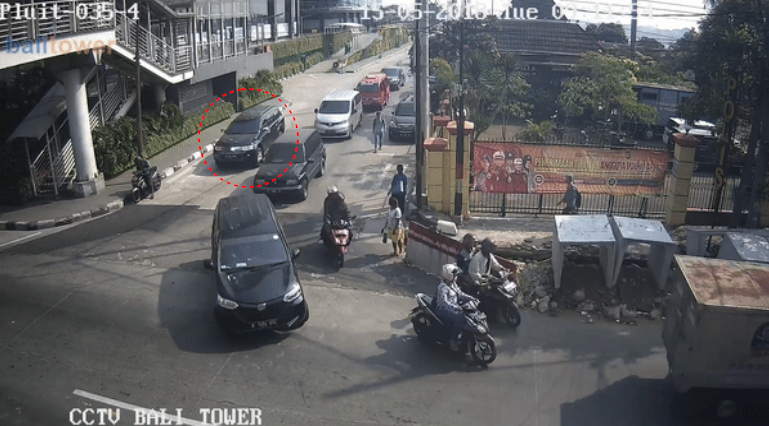}
    \end{subfigure}
    ~ 
    \begin{subfigure}[]{0.45\textwidth}
        \includegraphics[width=\textwidth, height = 2.5 cm]{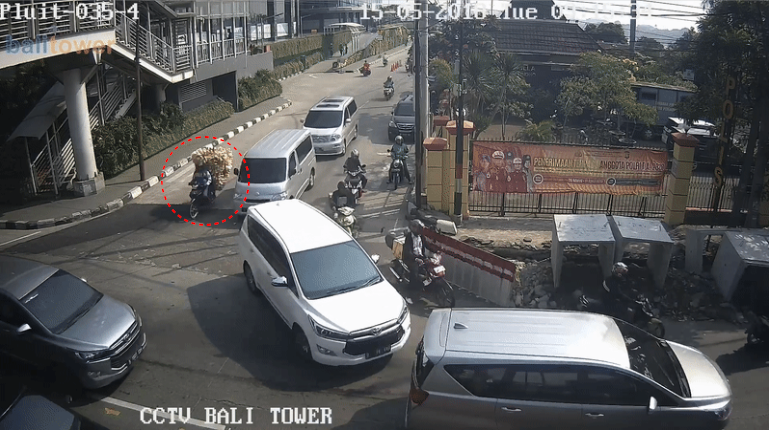}
    \end{subfigure}
    ~
    \begin{subfigure}[]{0.45\textwidth}
        \includegraphics[width=\textwidth, height = 2.5 cm]{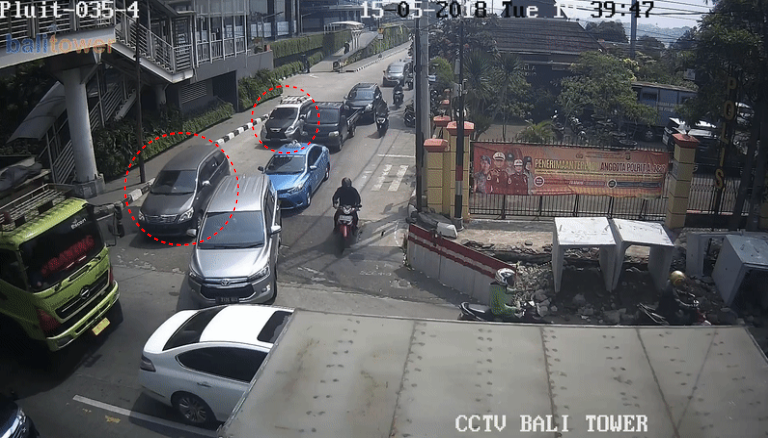}
    \end{subfigure}
    \caption{Examples of driving on the wrong side of the road found by our pipeline.}\label{fig:animals}
\end{figure}

Combining these methods, we can answer questions such as, ``Is this vehicle traveling on the wrong side of the road?'' or ``Is this motorcycle illegally parked on a sidewalk?'' Figure~\ref{fig:animals} shows one example of this. In this case, our system flagged four instances of a car moving in the wrong lane within a three-day span. In fact, three of these instances occurred in the same 2-hour period. One can imagine the utility such a system could provide, as an analyst can quickly identify that this intersection sees problematic behavior at particular days and times. This insight can then be used to inform interventions such as building a traffic light or median, or deploying a traffic steward at a busy time of day.

\section{Evaluation}

We evaluated object detection, classification, and motion detection by comparing our model outputs to the ground truth. 

For detection, we measure precision and recall. In this case, recall is the proportion of objects of interest which are correctly identified as objects, regardless of the predicted class. Precision is the proportion of detections which are true objects of interest. To evaluate these metrics for this specific problem, we go through all boxes predicted by our model in decreasing order of confidence. Ideally, the box drawn by the model will exactly align with the box drawn by the human, but in practice there will be differences. We used an ``Intersection Over Union'' (IOU) approach to determine whether two boxes were the same. If the IOU between the predicted and a true box are above our chosen threshold, we take those boxes to refer to the same object. Then we check if the predicted class is the same as the true class.  

\begin{figure}[ht]
    \centering
    ~ 
    \begin{subfigure}[t]{0.4\textwidth}
        \includegraphics[width=\textwidth, height = 4.5cm]{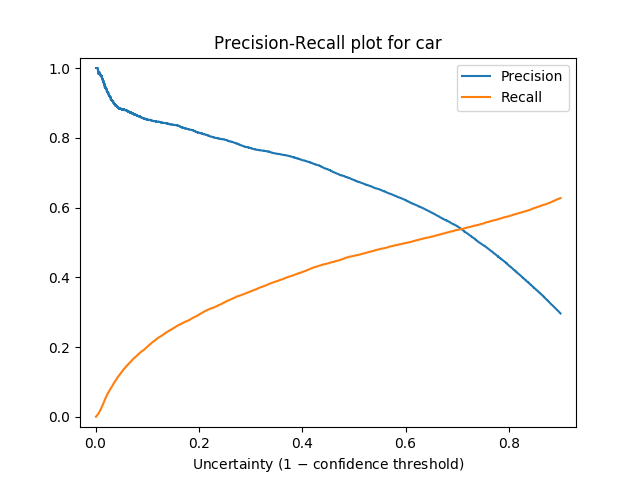}
        \caption{Example Precision-Recall plot for car detection and classification throughout all annotated videos, with IOU threshold set to 0.25. Note that with a confidence threshold of 0.5, we label 50\% of the cars in our labeled videos correctly, with 70\% precision. These results can be improved for instance by selecting only cameras with clearer perspectives.}
        \label{fig:pr_thresh_car}
    \end{subfigure}\hfill
    \begin{subfigure}[t]{0.55\textwidth}
        \includegraphics[width=\textwidth, height = 5cm]{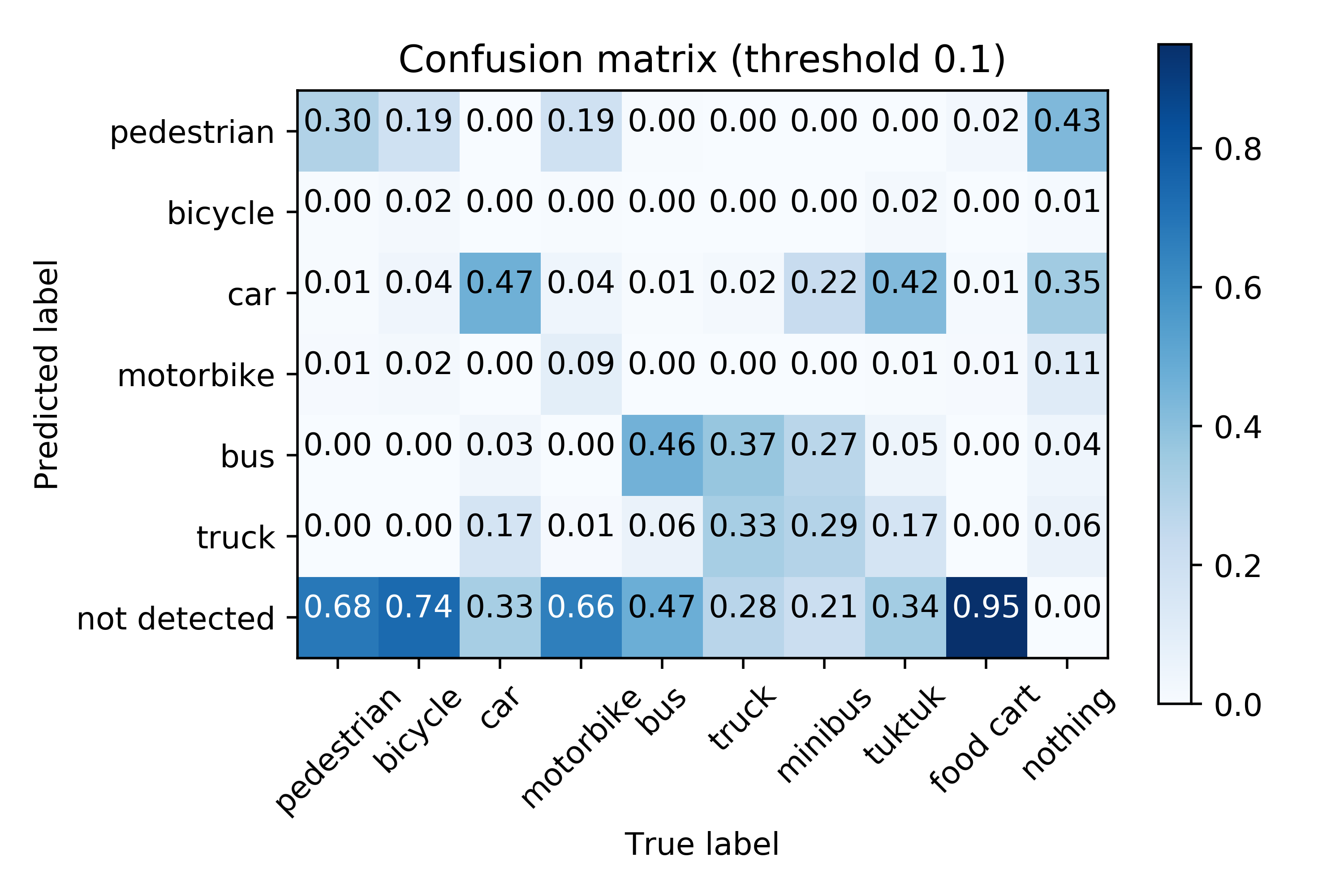}
        \caption{Confusion matrix, normalized so columns sum to 1. The objectness threshold is 0.4, while the label threshold is 0.1. One important finding is that while our algorithm does not have labels specific to Jakarta such as tuk-tuks and minibuses, we detect them roughly as well as cars or trucks, labeling them as cars, buses or trucks. If one is only interested in the behavior of the vehicles and not so much on the identification, it might not be necessary to fine-tune the model to detect these categories specifically. Other performance issues should be addressed by testing different object detection algorithms.}
        \label{fig:cf_matrix}
    \end{subfigure}
    \caption{Evaluation of object detection and classification.}
\end{figure}

There will be many parameters in the models that can be changed, including the IOU threshold considered in evaluation. We took a similar approach to classification evaluation, and allowed the pipeline to vary thresholds for objectness and labeling. Doing a grid search across these thresholds yielded several confusion matrices. An example result from this evaluation can be seen in Figure~\ref{fig:cf_matrix}. For this choice of object threshold, a large proportion of objects are not detected and those that are show varying levels of accuracy. 

We point out that some class confusion may be immaterial to the partners' ultimate intervention decisions. For example, consider an intersection where illegal left turns can pose a risk to opposing traffic. City planners would benefit from knowing whether large, heavy vehicles are making such illegal left turns, but it may be less important to distinguish between buses and trucks, as both pose comparable risks. However, confusing a motorbike for a pedestrian may contribute to a misconceived understanding of ground truth which will lead to improper policy decisions. Indeed, in our own evaluation, we saw noticed that tuk-tuks and mini-buses (which are Jakarta-specific and not in the YOLOv3 training set) were generally correctly characterized as something close to a car, but motorcycles and bicycles were frequently confused. Therefore this implementation of object detection and classification needs improvement. Thankfully this is a well studied problem in image analysis and there are several options for doing so. One of the chief goals in the near future is to experiment with various alternative models to find one better suited for the Jakarta context.


We also evaluated motion detection. For optimal settings, the average angle between detected and true motion is 11.0º. We can also use motion detection to effectively find vehicles moving in the wrong direction in a particular road, as we show in Figure~\ref{fig:animals}.

\section{Conclusion and Next Steps}

We set out to provide value to Jakarta by demonstrating what current technology could achieve as they prepare to deploy a complete video analysis system. Our ultimate aspiration is that this project will provide a template for how data scientists, local governments, NGOs, and the private sector can come together to advance urban policy. More than half the world's population now lives in cities, and this number continues to grow. As cities around the world grapple with rapidly growing populations, giving them the tools necessary to effectively manage transit will help guarantee their future safety and prosperity.

Starting in 2017, JSC has been building a big data infrastructure, and is looking to integrate the pipeline into its existing systems. This will consist of two distinct phases. The first phase will test the system by deploying it on a sample of CCTV cameras in Jakarta. Assuming these tests are successful, the system will then be integrated into every CCTV camera in the city. The second phase will focus on creating information systems that ensure that the results are disseminated to the relevant agencies in the Government of Jakarta.

The first phase will identify several roads in Jakarta that represent two categories: problematic and safe. Roads will be categorized by mining traffic data in collaboration with the Jakarta Transport Authority. After classifying roads, JSC will deploy the system on a sample of the CCTVs that monitor problematic roads, and then record the output (e.g. the number of detected cars, motorcycles, etc.). JSC will then validate and tune the classification, motion, and segmentation models. Once the models are well calibrated and fully deployed in the initial sample, JSC, together with Jakarta Transport Authority, will gather and interpret results, and then formulate and implement interventions on problematic roads. The effects of these interventions will be monitored, so the interventions can be continuously updated accordingly.

In the second phase, JSC will connect all of its CCTVs to the system. They will perform validation and verification, and make necessary model improvements. Once the models output information correctly and seamlessly, JSC will build information systems that support reports and/or a dashboard that will help various agencies in the Government of Jakarta understand model outputs, and hopefully improve decision making.

We hope our work illuminates the promise of using data to improve urban life around the globe.
The code developed for this project \href{https://github.com/dssg/jakarta_smart_city_traffic_safety_public/}{is available on GitHub} and we hope it proves valuable to anyone who wishes to develop or deploy a similar system and methods.

\bibliographystyle{unsrt}
\bibliography{jakarta_bib}

\end{document}